\documentstyle[aps,prl,multicol,epsf]{revtex}

\begin{document}
\def\CC{{\rm\kern.24em \vrule width.04em height1.46ex depth-.07ex
\kern-.30em C}}
\def\P{{\rm I\kern-.25em P}}
\def\RR{{\rm
         \vrule width.04em height1.58ex depth-.0ex
         \kern-.04em R}}

\draft
\title{  Stabilizing  Quantum Information}
\author{Paolo Zanardi }

\address{
 Istituto Nazionale per la Fisica della Materia (INFM) \\
 Institute for Scientific Interchange  Foundation, \\Villa Gualino,
Viale Settimio Severo 65, I-10133 Torino, Italy\\
}
\date{\today}
\maketitle

\begin{abstract}
The  dynamical-algebraic structure  underlying   all the schemes for quantum information
stabilization is argued to be fully contained in
the reducibility of  the operator algebra
describing the interaction with the environment 
of   the coding quantum system.
This property amounts to the existence of a non-trivial
group of symmetries for the global dynamics.
We  provide a unified framework
which allows us to build systematically new classes of  error correcting
codes and noiseless subsystems.
It is shown how by using  symmetrization  strategies one
can artificially produce noiseless subsystems
supporting universal quantum computation.

\end{abstract}
\pacs{PACS numbers: 03.67.Lx, 03.65.Fd}

\begin{multicols}{2}
Defending  quantum coherence  of a processing device 
against
the environmental interactions
is a vital goal for
 any foreseeable practical application of
Quantum Information and Quantum Computation theory \cite{QC}.
So far  basically three kind of strategies  have been devised
in order to satisfy such a crucial requirement:
a) Error Correcting Codes (ECC) \cite{ERR} 
which, in analogy with classical information theory, stabilize actively
quantum information by using redundant encoding 
and measurements; b) Error Avoiding Codes (EA) \cite{EA}
pursue a passive stabilization 
by exploiting symmetry properties  of the environment-induced noise 
for suitable redundant encoding;
c)  Noise suppression schemes \cite{SUPPR}
 in which, with no redundant encoding,
the decoherence-inducing interactions are averaged away
by properly tailored external ``pulses'' frequently iterated.
In this paper we shall show how all these schemes 
derive conceptually from  a common dynamical-algebraic framework.
The key notion to shed light on this underlying  structure
is that of {\em Noiseless Subsystem} (NS)  introduced  by Knill et al in ref. \cite{KLV}.
In this paper we shall discuss  how 
one can  analyze in a unified fashion in terms of purely algebraic data all the possible
strategies for quantum information stabilization. As a by product
a family of generalized ECC's will be introduced.
We shall provide 
abstract characterization of quantum evolutions  that
support NS's, and show how to obtain them by symmetrization procedures \cite{SYMM}.
Application to a realistic model of decoherence is given as well.

Let $S$ be an open quantum system, with (finite-dimensional) state-space ${\cal H},$
and self-Hamiltonian $H_S,$ coupled to its environment  through the hamiltonian
$H_I= \sum_\alpha S_\alpha\otimes B_\alpha,
$
where the $S_\alpha$'s  ($B_\alpha$'s) are system (environment) operators.
The unital associative algebra $\cal A$ 
closed under hermitian conjugation $S\mapsto S^\dagger,$
generated by the  $S_\alpha$'s \cite{uni}
and $H_S$ will be referred to as the {\em interaction} algebra. 
[We shall sometime identify $H_S$ with one of the $S_\alpha$'s
and discard the closed  case $H_I=0.$ 
]
The algebraic approach
used in this paper is not restricted to a Hamiltonian  description of
the dynamics.
Alternatively
 the dynamics of $S$ can be  described by
I) A Markovian master equation i.e., $\dot\rho=-i[H_S,\,\rho]+
1/2 \sum_\mu\lambda_\mu \{ [L_\mu\,\rho,\,L_\mu^\dagger]+  [L_\mu,\,\rho\,L_\mu^\dagger]\},$
for the density matrix $\rho.$
II) A finite time trace-preserving CP map $\rho\mapsto {\cal E}_t(\rho):=\sum_i e_i\,\rho\,e_i^
\dagger,\,
(\sum_i e_i^\dagger\,e_i=\openone.$)
In the first case the relevant interaction algebra is the one generated by $H_S$ and the
Lindblad operators $L_\mu.$
In the latter case $\cal A$ is generated by the ``error'' operators $e_i$'s.

In general $\cal A$ is a {\em reducible} $^\dagger$-closed 
subalgebra of the algebra $\mbox{End}({\cal H})$ of all the linear
operators over $\cal H.$
This implies that $\cal A$  can be written as a 
direct sum of $d_J\times d_J$ (complex) matrix algebras each  one of which appears
with a multiplicity $n_J$ \cite{CURE}
\begin{equation}
{\cal A}\cong\oplus_{J\in{\cal J}} \openone_{n_J}\otimes M(d_J,\,\CC).
\label{alg-split}
\end{equation}
where ${\cal J}$ is suitable finite set labelling the 
irreducible components of $\cal A.$
The  associated state-space decomposition  reads
\begin{equation}
{\cal H}\cong \oplus_{J\in{\cal J}} \CC^{n_J}\otimes \CC^{d_J}.
\label{split}
\end{equation}
These decompositions encode all  information about 
the possible quantum stabilization  strategies.

In ref. \cite{KLV} the authors observed that in view of relation
(\ref{alg-split}) each factor $\CC^{n_J}$  in eq. (\ref{split})
corresponds to a sort of effective subsystem of $S$  coupled to the  environment
in a state independent way.
Such subsystems are then referred to as {\em noiseless}.
In particular one gets a {\em noiseless code} i.e., a decoherence-free subspace, 
${\cal C}\subset {\cal H}$ when in equation (\ref{split}) 
there appear one-dimensional irreps $J_0$ with multiplicity greater than one 
${\cal C}\cong \CC^{n_{J_0}}\otimes \CC$ \cite{EA}.
The physical idea is very simple: one wants to identify a subspace of states
that corresponds to a multi-partite system in which one of the subsystems is coupled
with  the environment in such a way that quantum information cannot be extracted  from it.

We define the commutant ${\cal A}^\prime$ in End$({\cal H})$ of $\cal A$
by $ 
{\cal A}^\prime:=\{ X\,|\, [X,\,{\cal A}]=0\}.
$
From equation (\ref{alg-split}) it is clear that
the existence of  a NS is equivalent to 
${\cal A}^\prime
\cong
\oplus_{J\in{\cal J}}  M(n_J,\,\CC)\otimes \openone_{d_J}
\neq \CC\openone:=\{\lambda\,\openone\,|\,\lambda\in\CC\}.$
For a NS to be  relevant for  quantum encoding it must be at least
two-dimensional i.e., max$_J \{n_J\}\ge 2.$ This amount to have a {\em non-commutative} ${\cal A}^\prime.$
An interaction algebra satisfying the above condition will be called  NS-supporting.  
Of course when dim ${\cal A}^\prime=\sum_J n_J^2=1$  one is in the irreducible case 
[$|{\cal J}|=n_J=1$]
in which no NS exist.

In order to understand in what sense the NS's can be regarded as subsystems let us consider
the  projectors $Q_J:= \openone_{n_J}\otimes \openone_{d_J}\in {\cal A}\cap {\cal A}^\prime;$
they correspond to conserved   observables  that  constraint
the accessible state-space to one of the summands in eq. (\ref{split}) i.e., $Q_J\,{\cal H}$.
The identification of a bipartite structure stems from
the fact that on the ``superselection sector'' 
$Q_J\,{\cal H}$
the full operator algebra is isomorphic to 
 ${\cal A}\,{\cal A}^\prime\cong{\cal A}\otimes{\cal A}^\prime$ \cite{bipartite}.
The   duality ${\cal A}\mapsto {\cal A}^\prime,$ 
that will be repeatedly used later, is in this  sense  the algebraic
ground for the notion of subsystem.

An important special case is when  $\{S_\alpha\}$ is a commuting set of hermitian operators.
Then $\cal A$ is an abelian algebra and Eq. (\ref{split}) [with $d_J=1$]
is the decomposition of the state-space according the joint eigenspaces  
of the $S_\alpha$'s. The pointer basis \cite{ZUR} discussed in relation to the so-called
environment-induced superselection  is  nothing but an orthonormal 
basis associated to the resolution (\ref{split}).
The   NS's  provide
the natural non-commutative generalization of the pointer basis.
One might conjecture
that, for any initial preparation $\rho,$  a relation like $\lim_{t\to\infty} {\cal E}_t(\rho)
\in {\cal A}^\prime {\cal A}\cong \oplus_J M(n_J,\,\CC)\otimes  M(d_J,\,\CC),$
holds at least approximately.
[$\{{\cal E}_t\}$ denotes the dynamical semi-group]:
The quantum coherence between the different $J$
blocks is destroyed.

The decomposition (\ref{alg-split}) leads to a straightforward generalization
of the notion of stabilizer ECC \cite{stab} and allows us to build 
a general setting in which {\em non-additive} quantum codes \cite{NON} can arise.
Let $|J\lambda\mu\rangle \, (J\in{\cal J},\,\lambda=1,\ldots,n_J;\,\mu=1,\ldots,d_J)$
be an orthonormal basis associated to the decomposition (\ref{alg-split}).
Let
 ${\cal H}^J_\mu:=\mbox{span}\{|J\lambda\mu\rangle\,|\,\lambda=1,\ldots,n_J\},$
and let ${\cal H}^J_\lambda$ be defined analogously.
Now we consider a CP-map description of the dynamics
(see point  II) in the introduction),
the interaction algebra $\cal A$ being
generated by  error operators. 
Next proposition   shows that to any  NS corresponds a family of ECC's
(for a similar one see theorem 6 in ref. \cite{KLV}).

{\bf{Proposition 1.}}
{\em
The   ${\cal H}^J_{\mu}$'s are  ECCs
for any subset of  errors in ${\cal A}.$
}

{\em Proof.}
If $e_i, e_j\in  {\cal A}$ then $e_i^\dagger e_j\in{\cal A}.$
From Eq. (\ref{alg-split}) and the general results on ECC's  
\cite{ERR} the following computation
now suffices
$
\langle J \lambda^\prime\mu | e^\dagger_i\,e_j |J\lambda\mu\rangle= 
\langle J \lambda^\prime\mu |\openone\otimes X_{ij}
|J\lambda\mu\rangle =\delta_{\lambda, \lambda^\prime} c^{ij}_{J,\mu}.
$ 
$\hfill\Box$

This kind of ECC's will be referred to as $\cal A$-{\em codes}.
The above result  
extends to    any error set $E$ such
that $\forall e_i, e_j\in E\Rightarrow e_i^\dagger e_j\in{\cal B}$
where $\cal B$ is an operator algebra for which (\ref{alg-split}) holds.
The    proof above should make clear that the ${\cal H}^J_\lambda$ are ${\cal A}^\prime$-codes.
One recovers the usual picture  by considering
a $N$-partite qubit system,
and an abelian subgroup $\cal G$ 
of the Pauli group ${\cal P}:=\{\openone,\,\sigma_x,\,\sigma_y,\,\sigma_z\}^{\otimes\,N}.$
Let us consider the  state-space decomposition (\ref{split}) associated to $\cal G.$
If $\cal G$ has $k<N$ generators 
then $|{\cal G}|=2^k,$ whereas from commutativity  it follows  $d_J=1$ and $|{\cal J}|=|{\cal G}|.$
Moreover one finds
$n_J=2^{N-k}:$ each of the $2^k$ joint eigenspaces of ${\cal G}$ (stabilizer code)
 encode $N-k$ logical qubits.
Therefore one has  ${\cal H}=\oplus_{J=1}^{2^k} \CC^{2^{N-k}}\otimes
\CC\cong  \CC^{2^{N-k}}\otimes  \CC^{2^k}.$
Now it is known \cite{ortho} that correctable errors (belonging to the Pauli group)
correspond   to 
elements   $e_i,\,e_j$ such that $e_i^\dagger e_j$ either belongs to $\cal G$ or {\em anticommutes}
with (at least) one element $\cal G.$
In particular the latter operators induce a non-trivial mixing of different eigenspaces
i.e., a non-trivial action on the $ \CC^{2^k}$ factor.
In both cases they belong to the   algebra  ${\cal B}=\openone_{2^{N-k}}\otimes M(2^{k},\,\CC).$ 
The $\CC^{2^k}$ factor  corresponds in the usual stabilizer construction
to the encoding of the error syndrome, i.e., it will be a bitstring
containing the eigenvalues of the stabilizer. The errors correspond to
operations on this factor.

An  example of this construction is given by considering
any noiseless code. 
In this case since ${\cal A}|_{\cal C}\cong \openone_{n_0}\otimes M(1,\,\CC)$
one finds $c^{i j}_{0,1}= c_i\,c_j,$ since this matrix is not full rank
a noiseless code is a degenerate ECC \cite{LID1}.

It is well-known 
that group-theoretical notions  play
a key role in the analysis of all the 
schemes 
so far devised
for quantum-noise control. This  is true
for the study of general NS-supporting dynamics as well.
Indeed the 
condition ${\cal A}^\prime \neq \CC\openone$
 implies the existence of a non-trivial
group of symmetries ${\cal G}\subset U\,{\cal A}^\prime.$
Conversely given a group $\cal G$ of unitary operators over $\cal H$
its commutant is a reducible subalgebra of End$({\cal H})$
closed under hermitian conjugation.
Loosely speaking the more symmetric a dynamics, the more
likely it is  NS-supporting.

Therefore one is naturally led
to consider the  action, via a representation $\rho$,
of a finite  order (or compact) group $ {\cal G}$  on a quantum state-space $\cal H.$
The  irrep decomposition for $\rho$ has the form of Eq. (\ref{split})
where now  
the  ${\cal J}$ labels a set of  ${\cal G}$-irreps $\rho_J$ (dim $\rho_J=d_J$).
Extending $\rho$ by linearity to the group algebra 
$\CC{\cal G}:= \oplus_{g\in{\cal G}} \CC |g\rangle$  one gets 
a decomposition like in eq (\ref{alg-split}). 
It is now easy to provides a sufficient condition for an interaction algebra
to be NS-supporting

{\bf{Proposition 2.}}
{\em
If ${\cal A}\subset \rho(\CC\,{\cal G})$
then the dynamics supports (at least) $|{\cal J}|$ NS's   
with dimensions $\{n_J(\rho)\}_{J\in{\cal J}}.$
}


When $\cal G$ is a compact group Prop. 2  holds by replacing $\rho(\CC\,{\cal G})$
with the associative algebra generated by $\tilde\rho( {\cal L}),$
where $\cal L$  denotes Lie algebra of $\cal G$ and $\tilde\rho$ its representation
associated to $\rho.$
An important instance of this latter case is given by collective decoherence 
that will be discussed later in a more detailed manner.     
It should be stressed that the condition of belonging to
a group algebra is always  satisfied: it is sufficient 
to consider any  group acting irreducibly over $\cal H,$
e.g., the Pauli group
 in $N$-partite qubit systems.
The non trivial assumption is the reducibility of $\rho,$
when this is not given one has, in order to achieve it,  to resort to 
physical procedures for modifying the system dynamics.

{\em Symmetrizing.}
Now we address the issue of the relation between NS-supporting dynamics and 
the quantum noise  suppression schemes
recently emerged as a third possible way to
defeat decoherence in quantum computers \cite{SUPPR}.
In references \cite{VIO1} and \cite{SYMM}
it was  discussed how one can  devise {\em physical} procedures,
involving iterated external pulses or measurements, whereby
a quantum  dynamics generated by $\cal A$ can be modified
to a dynamics generated by $\pi_\rho({\cal A}).$
Here the ``symmetrizing'' projector $\pi_\rho$ is given by \cite{SYMM}
$
 X\rightarrow\pi_\rho(X):={|{\cal G}|^{-1}}\sum_{g\in{\cal G}}
\rho_g\,X\,\rho_g^\dagger\in\rho(\CC{\cal G})^\prime.
$
If we are willing  to retain the system self-dynamics
(generated by $H_S$) and to get rid just of the unwanted interaction with  the environment
(the $S_\alpha$'s),  
then we have to  look for a  group ${\cal G}\subset U({\cal H}),$
such that 
i) $H_S\in \CC{\cal G}^\prime,$ ii) the interaction operators $S_\alpha$
transform according to non-trivial   irreps under the (adjoint) action of $\cal G.$
In this case, since $\pi_{\cal G}$ projects
on ${\cal G}$-invariant i.e., trivial irrep, sector of End$({\cal H})$, it can be shown that
  $\pi_{\cal G}(S_\alpha)=0:$
the decoherence-inducing interaction have been averaged away, then the effective dynamics in unitary.

To make a connection between 
 noise suppression schemes and NS's a  crucial remark is to
notice that  Prop. 2  holds even by replacing the group algebra with its commutant
and the $n_J$'s with the $d_J$'s.
Indeed, since the  $\cal G$-symmetrization belongs to  $\rho(\CC{\cal G})^\prime,$
one has 
 
{\bf{Proposition 3.}}
 {\em 
The  ${\cal G}$-symmetrization of $\cal A$ supports (at least) $|{\cal J}|$ NS's  
with dimensions $\{d_J(\rho)\}_{J\in{\cal J}}.$}

The simplest instance of this result is given by $S$ being a $N$-level system
and ${\cal G}$ a finite group  that acts irreducibly over ${\cal H}$ e.g., an error generating
group \cite{KNI}.  Any ${\cal G}$-symmetrized interaction algebra is then proportional to the identity:
the whole space is a NS.
This situation corresponds to the  decoupling scheme analyzed in \cite{VIO}.

Next proposition straightforwardly generalizes a result of ref. \cite{Z}.
The key mathematical observations are 
i) The Lie algebra spanned 
by a {\em generic} couple of hermitian operators $H_1,\, H_2$ is the full u$({\cal H});$
ii)  the unitary group $U{\cal A}^\prime$ of the commutant
restricted to one of the summands  in eq. (\ref{split}) provide the full  
unitary group over the associated NS.
From i) it follows that, if one is physically   able to switch on and off $H_1$ and $H_2,$
any unitary transformation can be generated with arbitrary  accuracy \cite{SETH}.
More specifically, in view of ii), if one starts from Hamiltonians in ${\cal A}^\prime$ 
any unitary transformation over the NS can be (approximately) obtained.
Finally if such Hamiltonians are not available to the experimenter from the outset,
they can, in principle, be obtained from a generic i.e., not invariant, pair of Hamiltonians 
by a  symmetrization procedure \cite{Z}.
Formally:
 
{\bf{Proposition 4.}}
{\em
Given a  generic couple of Hamiltonians $\{H_i\}_{i=1}^2$ on the state-space of $S$
then their $\cal G$-symmetrizations 
$\{\pi_\rho(H_i)\}_{i=1}^2$ allow for universal quantum computation
over each of the NS's. 
}

This result   about universal quantum computation over a NS is just existential,  
nevertheless it is remarkable in that it 
shows how only a specific class of gates is required for generating arbitary
computations completely within the NS.
For practical purposes it is also important that the desired operations 
can be efficiently enacted in terms of physical interactions i.e., one- and two-body couplings.
This requirements must be checked case by case in that they do not follow from
Prop. 4.  
Constructive results for the case of collective decoherence
have been  recently found in ref. \cite{LID2},
in which it is shown  how to achieve universal computations
by resorting to  exchange Hamiltonians only.
More in general  it is likely the schemes with fast switching on and off
of Hamiltonians discussed  in ref. \cite{VIO} for  control of decoupled systems
will turn useful for achieving universal and efficient quantum computation over a NS.

{\em Collective Decoherence.}
Now we discuss the case of collective decoherence
when  a multi-partite quantum system,
whose  degrees of freedom are used for information encoding/processing,
is coupled symmetrically with a common environment.
This is the paradigmatic case for EA strategies \cite{EA}
[as opposed to ECC in which noise acts independently on each subsystem].
We shall show that it provides a  setting for 
NS's as well. Here the (minimal) symmetry group is the 
 the symmetric group ${\cal S}_N$ swapping different subsystems.
It follows that the  (maximal) interaction algebra that one can consider
is given by the space  
of totally symmetric operators. 
In the following we shall specialize to many-qubit systems.
All the results straightforwardly extend to general $d$-level systems
coupled to the their environment by  $sl (d)$ interactions.

Let us consider a $N$-qubit system  ${\cal H}_N:= (\CC^2)^{\otimes\,N}.$
 Over ${\cal H}_N$ acts the group $SU(2)$ via the $N$-fold (tensor) power of the defining irrep
i.e., $U\mapsto  U^{\otimes\,N}.$
The associated representation of the Lie algebra $su(2)=\mbox{span}\{\sigma_\alpha\}_{\alpha=1}^3$
is given, with obvious notation,  by
 $\rho_N\colon\sigma_\alpha\mapsto S_\alpha:=\sum_{i=1}^N \sigma^{(i)}_\alpha.$
The associative algebra generated by $\rho_N(su(2))$ will be denoted by ${\cal A}_N.$
We recall that \cite{CORN}:
1) ${\cal A}_N$ coincides with the algebra of completely symmetric operators
over ${\cal H}_N;$
2)
the commutant ${\cal A}_N^\prime$ is the group algebra $ \nu (\CC {\cal S}_N),$
where $\nu$ is   the natural representation of the symmetric group ${\cal S}_N$  over ${\cal H}_N:$
$\nu(\pi) \otimes_{j=1}^N |j\rangle = \otimes_{j=1}^N |\pi(j)\rangle,\,(\pi\in{\cal S}_N).$
From   $su(2)$ representation theory \cite{CORN} derives the following

{\bf{Proposition 5.}}
{\em
${\cal A}_N$ 
supports NS with dimensions 
$ 
n_J={[(2\,J+1)\,N!]}/{[(N/2+J+1)!\,(N/2-J)!]} 
$
where $J$ runs from $0$ ($1/2$) for $N$ even (odd).}

If in Prop. 5 ${\cal A}_N$ is replaced  by  its commutant, 
the above result holds with $n_J=2\,J+1.$
Moreover 
from Prop. 1 it is clear  that collective decoherence allows for
${\cal A}_N$-codes as well.
For example, let  us consider 
$N=3$ qubits.
One has $(\CC^2)^{\otimes\,3}\cong \CC\otimes \CC^4+ \CC^2\otimes\CC^2.$
The last term  can be  written as 
$\mbox{span}\{|\psi_\beta^\alpha\rangle\}_{\alpha\beta=1}^2
$
where
$|\psi_1^1\rangle =2^{-1/2}(|010\rangle-|100\rangle),\,
|\psi_2^1\rangle =2^{-1/2}(|011\rangle-|101\rangle)$
and
$|\psi_1^2\rangle ={2}/{\sqrt{6}}\,[1/2 (|010\rangle+|100\rangle)-|001\rangle],\,
|\psi_2^2\rangle ={2}/{\sqrt{6}}\,[|110\rangle-1/2(|011\rangle+|101\rangle)].$
One can check, for example, that  $|\psi_\beta^1\rangle$ and $|\psi_\beta^2\rangle$
( $|\psi_1^\alpha\rangle$ and $|\psi_2^\alpha\rangle$) span a two-dimensional ${\cal A}_3$-code
(${\cal A}^\prime_3$-code).  
Taking the trace with respect to the index $\alpha$ ($\beta$)
one gets the ${\cal A}_3^\prime$ (${\cal A}_3$) NS's.
Moreover the first term corresponds to a trivial four-dimensional
${\cal A}_3^\prime$  code.
Notice that any permutation
error
can be written as the product  of transpositions that in turn, 
in this representation, corresponds  to the so-called  exchange errors \cite{PAU}.

A weaker kind of collective decoherence is obtained  when the symmetry
group breaks down: 
${\cal S}_N  \rightarrow \prod_{c=1}^R {\cal S}_{n_c}, \,(\sum_{c=1}^R n_c=N).
$
The maximal $NS$-supporting interaction algebra is then isomorphic to  the tensor product
$\otimes_{c=1}^R {\cal A}_{n_c},$ for which the obvious extension of Prop. 5 holds:
NS's exist , given by all possible tensor products of the cluster NS's.
Physically this situation corresponds to  $R$ uncorrelated clusters of subsystems
such that within each cluster the condition of collective decoherence is fulfilled \cite{clust}.
As limiting cases one obtain   collective decoherence
and independent one in which no NS's exist

We finally comment on possible infinite-dimensional 
extensions of the ideas and results presented in this paper.
They would be relevant for  quantum computation with continuous variables \cite{CONT}.
The crucial observation in this respect  is that, 
adding a  suitable closure assumption on the interaction algebra,
a generalized form of the basic decompositions (\ref{split}) and (\ref{alg-split})
holds \cite{CURE} p. 9. It is then likely that, at least some of the constructions
of this paper  
can be properly reformulated in the continuous case.
This important issue will be addressed elsewhere; here we limit ourselves
to a very simple example that represents the continuous analog of collective decoherence
case previously discussed.
Let us consider $N$ copies of a continuous quantum system described by 
conjugate variables $x_j,\,p_l,\,( [x_j,\,p_l]=i \,\delta_{lj})$
coupled with a common environment only through the collective coordinates
$X:= \sum_{j=1}^N x_j,\,P:=  \sum_{j=1}^N p_j.$
This  assumption implies that the interaction Hamiltonian can be written as
$H_I= \sum_\alpha f_\alpha(X,\,P) \otimes B_\alpha,$
where the $f_\alpha$ are operator-valued functions generating the relevant
interaction algebra ${\cal A}_\infty$ 
and the $B_\alpha$'s
environment operators.
We define creation and annihilation  operators 
by $a_k^\pm:=1/\sqrt{2N}\sum_{j=1}^N \exp(i \frac{2\pi}{N} kj)(x_j\pm i\,p_j),\,(k=0,\ldots,N-1)$ 
one has End$({\cal H})\cong\otimes_k {\cal A}_k,$ where ${\cal A}_k$ denotes the algebra generated by
$\{a_k^\pm\}.$
One can check  
that  ${\cal A}_\infty\subset\openone_{k>0}\otimes{\cal A}_0.$ 
It follows that the factor of ${\cal H}$ corresponding to non-zero modes
realizes an infinite-dimensional NS.

{\em Conclusions.}
In this paper we faced the problem of stabilizing quantum information
against decoherence in a dynamical-algebraic fashion.
The analysis of the operator algebra $\cal A$ generated by 
interactions with the environment and the self-Hamiltonian 
of the information processing system provides the  general conceptual
framework.  The notion of noiseless subsystem \cite{KLV}  has
been shown to be the key tool for unveiling the common
structure  at the root of all the (quantum) error correction, error avoiding and
error suppression schemes discovered so far :
the reducibility of $\cal A$ provides sectors of the state-space
from which information cannot be extracted by unwanted interactions. 
New families of ECC's have been presented. We have described general symmetrizing 
strategies designed to  synthesize quantum  evolutions with  the desired capability
of supporting noiseless subsystems. 
The  general ideas have been exemplified by the  collective decoherence 
case.
In our opinion, the overall emerging picture  is conceptually quite satisfactory
in that, on the one hand it allows us to clarify the strict mutual relations
between apparently different techniques; on the other hand, in view of its generality, 
it is likely to open new ways to  practical realization 
of noiseless quantum information processing.

 I thank M. Rasetti for useful discussions and critical reading of the manuscript.
Elsag (a Finmeccanica company) for financial support.

\end{multicols}


\begin{references}
\bibitem{uni} Given  ${\cal S}\subset\mbox{End}({\cal H}),$ the associative
unital $^\dagger$-closed algebra generated by ${\cal S}$ is the norm closure of the linear span
of the monomials built with operators in ${\cal S}\cup{\cal S}^\dagger$ plus the identity $\openone.$
\end{references}
\end{document}